\documentstyle[12pt]{article}
\newcommand{\be}{\begin{equation}}
\newcommand{\ee}{\end{equation}}
\newcommand{\bea}{\begin{eqnarray}}
\newcommand{\ena}{\end{eqnarray}}
\newcommand{\vs}[1]{\rule[- #1 mm]{0mm}{#1 mm}}

\newcommand{\gamgam}{$\gamma - \gamma\ $}
\newcommand{\epem}{$\ e^+ \ e^-\ $}
\newcommand{\rs}{${\sqrt s}\ $}
\newcommand{\rsee}{${\sqrt s_{e^+e^-}}$}
\newcommand{\pT}{${p_{T} }\ $}

\newcommand{\MSB}{\overline {MS}}

\newcommand{\alfapi}{{\alpha \over 2 \pi}}

\newcommand{\as}{\alpha_{s}(\mu)}

\newcommand{\asss}{\alpha_{s}^3(\mu)}
\newcommand{\alfspi}{ {\alpha_s(\mu) \over 2 \pi} }
\newcommand{\alfspis}{ \left( {\alpha_s(\mu) \over 2 \pi} \right) ^2 }

\newcommand{\lptm}{ \ln  {p^2_T \over M^2}  }
\newcommand{\lmupt}{ \ln  {\mu ^2 \over p^2_T}  }
\newcommand{\dsigee}{  {d \sigma^{e^+e^- \rightarrow jet}
            \over {d {\vec p}_T d \eta}}}
\newcommand{\dsigd}{  {d \sigma^{D} \over {d {\vec p}_T d \eta}}}
\newcommand{\dsigsf}{{d \sigma^{SF} \over {d {\vec p}_T d \eta}}}
\newcommand{\dsigdf}{{d \sigma^{DF} \over {d {\vec p}_T d \eta}}}
\newcommand{\dsiggg}{ {d \sigma^{\gamma \gamma \rightarrow jet}
            \over {d {\vec p}_T d \eta}}}
\newcommand{\dsigig}{{d \sigma^{i \gamma \rightarrow jet}
            \over {d {\vec p}_T d \eta}}}

\newcommand{\dsigqg}{{d \sigma^{q \gamma \rightarrow jet}
            \over {d {\vec p}_T d \eta}}}
\newcommand{\dsiggq}{{d \sigma^{\gamma q \rightarrow jet}
            \over {d {\vec p}_T d \eta}}}
\newcommand{\dsigjg}{{d \sigma^{j \gamma \rightarrow jet}
            \over {d {\vec p}_T d \eta}}}
\newcommand{\dsiggj}{{d \sigma^{\gamma j \rightarrow jet}
            \over {d {\vec p}_T d \eta}}}
\newcommand{\dsigij}{{d \sigma^{i j \rightarrow jet}
            \over {d {\vec p}_T d \eta}}}
\newcommand{\dsigkj}{{d \sigma^{k j \rightarrow jet}
            \over {d {\vec p}_T d \eta}}}
\newcommand{\dsigil}{{d \sigma^{i l \rightarrow jet}
            \over {d {\vec p}_T d \eta}}}

\newcommand{\NP}[1]{Nucl.\ Phys.\ {\bf #1}}
\newcommand{\PL}[1]{Phys.\ Lett.\ {\bf #1}}

\newcommand{\PR}[1]{Phys.\ Rev.\ {\bf #1}}
\newcommand{\PRL}[1]{Phys.\ Rev.\ Lett.\ {\bf #1}}
\newcommand{\ZPH}[1]{Z.\ Phys.\  {\bf #1}}

\begin{document}

\renewcommand{\thefootnote}{\fnsymbol{footnote}}
\newpage
\setcounter{page}{0}

\rightline{KEK-CP-11}
\rightline{KEK Preprint 93-180} 
\rightline{LAPP-TH-436/93}
\rightline{LPTHE-ORSAY 93/47}
\rightline{December 1993}

\vs{9}

\begin{center}
{\Large {\bf{Jets in photon-photon collisions: from TRISTAN to J/N-LC  }}} \\
\vspace{0.7 cm}
{\large P. Aurenche, J.-Ph. Guillet} \\
{\em Laboratoire de Physique Th\'eorique ENS{\large{\em L}}APP
\footnote{URA 14-36 du
CNRS, associ\'ee \`a l'Ecole Normale Sup\'erieure de Lyon, et au
Laboratoire d'Annecy-le-Vieux de Physique des Particules.} $-$ Groupe
d'Annecy\\
LAPP, IN2P3-CNRS, B.P. 110, F-74941 Annecy-le-Vieux Cedex, France}
\\[0.7cm]
{\large M. Fontannaz} \\
{\em Laboratoire de Physique Th\'eorique et Hautes Energies
\footnote {URA 63 du CNRS, associ\'ee \`a l'Universit\'e de Paris XI.} \\
Universit\'e de Paris XI, b\^atiment 211, F-91405 Orsay Cedex, France}
\\[0.7cm]
{\large Y. Shimizu, J. Fujimoto} \\
{\em Physics Department, KEK \\
Tsukuba, Ibaraki 305, Japan} \\[0.8cm]
{\large K. Kato} \\
{\em Physics Department, Kogakuin University \\
Shinjuku, Tokyo 160, Japan} \\[0.8cm]
\end{center}
\vs{4}

\centerline{ \bf{Abstract}}
\vs{3}

We study jet production in photon-photon reactions at the next-to-leading
logarithm accuracy. Special emphasis is placed on the discussion of the
theoretical uncertainties and on the role of the hadronic component of
the photon structure function.
We also discuss subtleties in the quasi-real photon spectrum when the
photon is involved in a large transverse momentum reaction. The phenomenology
at TRISTAN energies is discussed and predictions are made for LEP and a $1\ TeV$
\epem collider.


\newpage

\section{ Introduction   }

\indent

  Very recently, several experimental collaborations have published results
concerning large transverse momenta phenomena in \gamgam collisions where the
incoming photons are real or quasi-real. In particular, the AMY \cite{1}
 and TOPAZ \cite{2}
collaborations have measured the high transverse momentum jet rates at \rsee
$=\ 58\ GeV$. At a lower energy (\rsee $=\ 29\ GeV$) a related observable,
namely the large \pT particle spectrum, has been measured at the
MARK II \cite{3} detector. Theoretical work on this subject started long
ago \cite{hist},\cite{7}. Here for the first time,  we discuss the QCD
predictions, at the
next-to-leading logarithm accuracy, of the production of jets in \gamgam
collisions. This study seems appropriate at this time
since the above data are rather
precise and, furthermore, data at higher energies are coming from LEP \cite{3a}.
It is well known that the rate of \gamgam events increases with energy in
contrast to the rate of \epem annihilation into a virtual photon in the
s-channel and therefore \gamgam large \pT phenomena will provide an increasingly
important background in the search of new phenomena at very high energies.

  A priori, hard physics in \gamgam collisions should be very simple since the
photon couples directly to the partons involved in the hard momentum transfer
process: it is indeed so in the kinematical region where the transverse
momentum is not too small compared to the initial energy. However, in the
intermediate \pT region the anomalous photon component \cite{4,5} introduces a
new contribution where the photon couples through its perturbative quark and
gluon contents. The problem is further complicated by the non-perturbative
hadronic component of the photon which is often described
in the vector meson dominance model (VDM). 
The (perturbative) anomalous and the (non-perturbative)
hadronic components can be dealt with by introducing the scale dependent
photon structure function very much in the same way one introduces a
hadron structure function, with the difference that the former is predicted by
QCD at very large scales. Consequently \gamgam physics becomes more complicated
than hadron-hadron collision physics since the photon interacts either
point-like or via its structure function.

   In the following we treat all these processes at the next-to-leading
logarithmic accuracy i.e. we include higher order corrections both in the
structure function and the hard subprocesses in a consistent way. In the
literature, the interaction of the photon through its partonic constituents is
often referred to as the ``resolved" photon interaction \cite{7}: if this
terminology may be  appropriate at the leading-logarithmic level its becomes
inadequate at the next level of accuracy since the non-leading part
of the ``resolved" photon
contributes to the higher order corrections of the direct photon process.

   Information about the photon structure function is obtained from different
experiments: the quark distribution is rather well constrained from
deep-inelastic scattering of a real photon which is, on the other hand,
not very sensitive to the gluon distribution in the photon. Preliminary
results at HERA \cite{8,9} have however shown that it is possible to constrain
the gluon distribution at rather
small values of the $x$ variable. The same can be said of \gamgam experiments
at TRISTAN and LEP although in a somewhat higher $x$ range because the \pT/\rs
values probed are larger than at HERA. Thus, by collecting experimental
information from all these types of reactions a good determination
of the photon structure function should be possible. In particular, the
non-perturbative gluon content should soon become under control.

  In the next section we set up the formalism and define the differential jet
cross section at the next-to-leading logarithmic level. We then make a
detailed study of the theoretical ambiguities (choice of renormalization and
factorization scales)
covering TRISTAN energies to \rsee $=\ 1\ TeV$. Before making a
comparison to the recent TOPAZ results we consider several problems related to
the approximations involved in the parametrization of the quasi-real photon
flux. In particular, we discuss the accuracy of the
Weizs\"acker-Williams \cite{9a} approximation to describe large \pT phenomena
under specific experimental tagging conditions. An effective spectrum is
introduced which is smaller than the usual one. Then we argue that the
virtuality of the photons in the anti-tag conditions used by TOPAZ are such that
form-factor effects have to be taken into account and this further reduce the
\gamgam collision rate. Finally typical predictions for LEP and a $1\ TeV$
collider are given.

\indent

\section{  The formalism }

\indent

   Following the previous discussion, one can naturally separate the single
inclusive jet cross section at large \pT according to
the type of interaction of the incoming photons in the hard sub-process. If
both photons couple directly to the quarks involved in the hard scattering one
defines the ``direct" cross section (fig. 1)
\bea
\dsigd (R) =  \dsiggg \ + \ \alfspi K^{D} (R;M) .
\label{eq:dir}
\ena
If one photon only interacts via the quark or the gluon component in its
structure function we define (fig. 2)
\bea
\dsigsf (R) &=& \sum_{i=q,g} \int dx_{1} F_{i/\gamma}(x_{1},M)
\nonumber \\
&\ & \ \ \ \ \ \alfspi \left( \dsigig + \ \alfspi K^{SF}_{i \gamma} (R;M,
\mu)
 \right) \nonumber \\
 &+& \sum_{j=q,g} \int dx_{2} F_{j/\gamma}(x_{2},M)
 \nonumber \\
&\ & \ \ \ \ \ \alfspi \left( \dsiggj + \ \alfspi K^{SF}_{\gamma j} (R;M,
\mu)
 \right)
\label{eq:sf}
\ena
Finally, if both photons interact via their structure functions we have (fig.
3)
\bea
\dsigdf (R) &=& \sum_{i,j=q,g} \int dx_{1} dx_{2}
 F_{i/\gamma}(x_{1},M) F_{j/\gamma}(x_{2},M) \nonumber \\
&\ & \ \ \ \ \ \ \ \
\alfspis \left( \dsigij \ + \ \alfspi K^{DF}_{ij} (R;M,\mu) \right)
\label{eq:df}
\ena
In writing the above equations we have not specified all the arguments of the
various functions but we have kept only those which are relevant for our
discussion. The variable $R$ is the usual parameter defining the size of the
detected jet with transverse momentum ${\vec p}_T$, pseudo-rapidity
$\eta$ and azimuthal angle $\phi$. When a jet is composed of two partons its
coordinates are defined as \cite{10a}, \cite{2}.
\bea
p_T  &=& \vert {\vec p}_{T_1} \vert + \vert  {\vec p}_{T_2} \vert \nonumber \\
\eta &=& {1 \over p_T} ( p_{T_1} \eta_1 + p_{T_2} \eta_ 2) \nonumber \\
\phi &=& {1 \over p_T} ( p_{T_1} \phi_1 + p_{T_2} \phi_ 2)
\label{eq:kin}
\ena
and the condition for the partons to  belong to the jet is
\bea
{\root \of {\vert \eta_i-\eta \vert^2 + \vert \phi_i-\phi \vert^2}} < R.
\label{eq:jet}
\ena
By convention, we have explicitly written the strong interaction coupling out
of the partonic cross sections of type
$d \sigma^{i j \rightarrow jet} / {d {\vec p}_T d \eta}$
and denoted $\mu$ the renormalization scale. The photon
structure function $F_{i/\gamma}(x,M)$ is the probability density to find a
parton of type $i$ carrying a fraction $x$ of the incoming photon when probed
at the factorization scale $M$. Combinations of structure functions satisfy
evolution equations beyond the leading order. If one defines, as usual,
singlet and non singlet quark distributions via:
\bea
\Sigma^\gamma &=& {\sum_{f=1}^{N_f}} F^+_{f/\gamma} \nonumber \\
F^+_{f/\gamma}&=&F_{f/\gamma} + F_{{\bar f} /\gamma} \nonumber \\
F^{NS}_{f/\gamma} &=& F^+_{f/\gamma} -{1 \over N_f} \Sigma^{\gamma}.
\label{eq:sing}
\ena
with $N_f$ the number of flavors, the evolution equations can be written
\bea
{d \Sigma^\gamma \over d \ln M^2}&=& k_q + P_{qq} \otimes \Sigma^\gamma
+ P_{qg} \otimes F_{g/\gamma} , \nonumber \\
{d F_{g/\gamma} \over d \ln M^2}&=& k_g + P_{gq} \otimes \Sigma^\gamma
+ P_{gg} \otimes F_{g/\gamma} , \nonumber \\
{d F^{NS}_{f/\gamma} \over d \ln M^2}&=& {1 \over N_f} \left(
{e^2_f \over < e^2 >} -1 \right) k_q + P_{NS} \otimes F^{NS}_{f/\gamma}.
\label{eq:evol}
\ena
The convolution symbol is defined by
\bea
P \otimes \Sigma^\gamma = \int^1_x dz\ P \left({x \over
z} \right) \ \Sigma^\gamma(z)
\ena
and the average charge squared is
\bea
< e^2 > = {1 \over N_f} \sum^{N_f}_{f=1} e^2_f .
\ena
We come back to the discussion of these equations shortly. In
eqs. (\ref{eq:dir})-(\ref{eq:df}) the functions $K^D,\ K^{SF}$ and $K^{DF}$
are the higher order corrections to the various cross sections. These functions
depend on the kinematical variables (not shown) as well as the
renormalization and factorization scales as indicated. The term $K^{DF}$
is the same as the higher correction term in the production of a large \pT
jet in hadron-hadron collisions: it has already been calculated \cite{11}.
As for $K^{SF}$ and $K^D$ they are obtained using the same technics as those
necessary to obtain $K^{DF}$ starting from the appropriate matrix elements which
can be found in \cite{12}. For a precise discussion on the factorized form
of eqs. (\ref{eq:dir})-(\ref{eq:df}) see \cite{13}.
It should be noted that eq. (\ref{eq:dir}) is of order $\as$ while
eq. (\ref{eq:df}) is of order $\asss$. However, since the structure functions
$F_{i/\gamma}(M)$ are known to be of order $\alpha_s^{-1}(M)$ all the
equations are in fact calculated consistently up to order $\alpha_s$ when
both $M$ and $\mu$ large.

\indent

       As we are now going to
discuss, a compensation in the unphysical factorization mass $M$ dependence
occurs between the different equations so that each individual equation is not
a physical quantity but only the sum of all three equations is the physical
observable. In other words, one should not try to associate, for example,
eq. (\ref{eq:sf}) to the one-resolved photon cross section since the
$M$ scale ambiguity in this quantity occurs at the order at
which the calculation is made.

 In eqs. (\ref{eq:evol}), the functions $P_{ij},\ P_{NS}$ and $k_i$
admit a perturbative expansion of the form
\bea
P_{ij} &=& {\alpha _s (M) \over 2 \pi} \left( P^{(0)}_{ij} +
         {\alpha _s (M) \over 2 \pi} P^{(1)}_{ij} \right),  \nonumber \\
k_{i} &=& {\alfapi} \left( k^{(0)}_{j} +
         {\alpha _s (M) \over 2 \pi} k^{(1)}_{j} \right)
\ena
The lowest order terms are the usual
Altarelli-Parisi splitting function (with $P^{(0)}_{qq}=P^{(0)}_{NS}$) appearing
in the evolution equation of the hadronic structure function. The inhomogeneous
coefficient $k^{(0)}_i$ are easily calculated and originate from the splitting
of the photon into a collinear quark-antiquark pair at the $0^{th}$ order in
QCD (the so-called box approximation). It is found
\bea
k^{(0)}_g(x) &=& 0                                              \nonumber \\
k^{(0)}_q(x) &=& 2 N_c N_f <e^2> \left( x^2+(1-x)^2 \right) \nonumber \\
         &=& 2 N_f P^{(0)}_{q \gamma}(x)   \nonumber \\
         &=& 2 N_f P^{(0)}_{{\bar q} \gamma}(x)      \label{eq:kgq}
\ena
The corresponding higher order terms
can be found in ref. \cite{14} with corrections first discussed
in ref. \cite{15}. To illustrate the scale compensation mechanism it is enough
to use eqs. (\ref{eq:evol}) to leading order. The discussion is further
simplified if we consider only one type of quark so that, for our purposes, the
evolution equations reduce to (where we have used for notational convenience
$P^{(0)}_{g \gamma} = k^{(0)}_g $)
\bea
{d F_{i/\gamma} \over d \ln M^2}&=& {\alfapi} P^{(0)}_{i \gamma}(x)
+ {\alpha _s (M) \over 2 \pi} P^{(0)}_{ij} \otimes F_{j/\gamma}
\label{eq:evolsimp}
\ena
where the label $i$ refers to a quark, an anti-quark or a gluon.
The higher order calculation yields the following factorization and
renormalization mass scale dependence for the correction terms:
\bea
K^D &=& \alfapi P^{(0)}_{q \gamma} \otimes \left[ \dsigqg + \dsiggq \right]
\lptm + \kappa^D                                                \nonumber \\
K^{SF}_{i \gamma}&=&\left[ \alfapi P^{(0)}_{j \gamma} \otimes \dsigij
+ P^{(0)}_{ij} \otimes \dsigjg \right] \lptm + b \dsigig \lmupt
+ \kappa^{SF}_{i \gamma}                                        \nonumber \\
K^{DF}_{ij}&=& \left[ P^{(0)}_{ik} \otimes \dsigkj + P^{(0)}_{jl} \otimes
\dsigil \right] \lptm
            + 2 b \dsigij \lmupt + \kappa^{DF}_{ij}
\label{eq:compens}
\ena
where the numerical coefficient $b$ controls the evolution of the strong
coupling at the leading order
($ 2 \pi d \alpha_s(\mu) / d \ln (\mu^2) = b \alpha_s^2(\mu)$).
A change in the factorization scale $M$ of the structure function
$F_{i/\gamma}$ induces, according to eq. (\ref{eq:evolsimp}), an inhomogeneous
variation controlled by $P^{(0)}_{i \gamma}$ and an homogeneous one proportional
to $\alpha_s P^{(0)}_{ij}$. The induced variation on
$ {d \sigma^{SF} / {d {\vec p}_T d \eta}}$ is then compensated, on the one
hand, by the term in $K^D$ proportional to $\ln (p^2_T / M^2)$ as far as the
inhomogeneous piece is
concerned and, on the other hand, by the terms in $K^{SF}$ containing
$P^{(0)}_{ij}$
as far as the homogeneous variation is concerned. The remaining terms in $K^{SF}$
linear in $\ln (p^2_T / M^2)$ serve to balance the inhomogeneous variation of
$ {d \sigma^{DF} / {d {\vec p}_T d \eta}}$ whereas the homogeneous variation is
compensated by the $M$ dependent terms of $K^{DF}$. This can be checked
explicitly by using eq. (\ref{eq:evolsimp}) to study the variation of
eqs. (\ref{eq:dir})-(\ref{eq:df}) under a change of $M$,
keeping the renormalization scale $\mu$ fixed.
As a consequence it is obvious that the sum of the $D,\ SF$
and $DF$ processes have a physical meaning at the next-to-leading logarithm
level since, for this combination, the factorization scale ambiguity is removed
at the order at which the calculation is performed.

The effect of a variation of the scale $\mu$ is more traditional since the
compensation mechanism takes place between the lower order and the higher
order terms of the $SF$ and $DF$ cross sections individually exactly as it is
the case for purely hadronic reactions. There is no
compensation in $d \sigma^D / {d {\vec p}_T d \eta}$ but the ambiguity appears
at the next-to-next-to leading logarithm level.

        We finally need to specify the functions $F_{i/\gamma}$,  i.e. the
parton distributions in the photon which appear in eqs.
(\ref{eq:sf})-(\ref{eq:df}).
These functions are discussed in great details in \cite{13} and we will not
dwell on this point here. We just recall that we use the
$\MSB$ convention for the structure functions as well as the
higher order terms. Furthermore, these functions which are solutions of eqs.
(\ref{eq:evol}) are the linear combinations of a (perturbative) anomalous term
which is chosen to vanish below the scale $Q_0^2=.5\ GeV^2$ and
(non-perturbative)
hadronic term, related via the usual VDM assumptions to the vector meson
structure functions (identified in a first approximation with the pion structure
functions). These distributions are in good agreement with the data on
$F_2^\gamma$ from PLUTO \cite{16}, AMY \cite{17}, JADE \cite{18}.

    We illustrate the above discussion by plotting, for a few relevant examples
(see figs. 4 and 5),
the variation of the theoretical cross section under independent changes of the
scales $M$ and $\mu$. The figures are obtained using the
standard photon structure functions \cite{13} and after folding the
$\gamma - \gamma$ cross section with
the appropriate Weizs\"acker-Williams spectrum to reconstruct the
observable $e^+ e^- \rightarrow jet$ as is explained in detail in the
next section.
We show the result for TOPAZ , \rsee $=\ 58\ GeV$ and $p_T=\ 5.24\ GeV/c$, as
well as for \rsee $=\ 1\ TeV$ and $p_T=\ 50\ GeV/c$. The general features are
the following: as expected the leading logarithm predictions are monotonously
increasing when $M$ increases and $\mu$ decreases whereas the next-to-leading
logarithm predictions have, in general,  a maximum in $\mu$ at a fixed
factorization scale. Unfortunately, there is no optimum in the variable $M$ for
a fixed renormalization scale except at unrealistically low $\mu$. This is due
to the extremely rapid variation of the $DF$ term, eq. (\ref{eq:df}), which is
only partially compensated by the variation of the $SF$ term in the opposite
direction. As a result there does not exist a region of
stability
(saddle point or absolute maximum or minimum)
of the cross section under changes of scale as is often found in
other perturbatively calculated QCD processes.
We nevertheless observe that the next-to-leading logarithm cross section is
considerably flatter than the leading logarithm one for $\mu > 1\ GeV$ in the
case of TOPAZ and $\mu > 3\ GeV$ in the case of J/N-LC.
We can also observe that
keeping the ratio $\mu / M$ fixed and changing the common scale both the
leading and the next-to-leading cross sections are surprisingly stable: this is
due to the compensation of the logarithmic dependence of the anomalous
component by the strong interaction coupling. The value of the cross sections
however depends on the value of the ratio of the two scales.
In the following, we will use the
standard choice $\mu=M$, keeping in mind the rather large uncertainty in the
theoretical predictions.
\indent

\section{  Phenomenology and comparison with TOPAZ data }

\indent

        The discussion in the previous section concerns jet production in
collisions between two real photons. The experimental situation is not so
simple since the measurements are done for jet production in {\epem}
via the exchange of two quasi-real photons. We discuss now the method and the
approximations involved in going from the theoretical results to the
experimental observables. The basic formula relating these two quantities
relies on the equivalent photon approximation or the Weizs\"acker-Williams
approximation \cite{9a} which factorizes the flux of quasi-real photons emitted
by the electron and the positron  from the interaction rate between the two
photons assumed to be real
\bea
{d \sigma^{e^+e^- \rightarrow jet} \over {d {\vec p}_T d \eta}} =
\int dz_{1} dz_{2}  F_{\gamma/ e}(z_{1},E) F_{\gamma/ e}(z_{2},E)
{d \sigma^{\gamma \gamma \rightarrow jet } \over {d {\vec p}_T d \eta}}
\label{eq:equiv}
\ena
where $E$ is the common energy of the electron and the positron. It has been
shown that the expression \cite{20}
\bea
F_{\gamma/ e}(z,E) &=& {\alpha \over \pi z} \left( \left( 1+(1-z)^2 \right)
\left( \ln{2 (1-z) E \over m_e } -\frac{1}{2}
\right) \right. \nonumber \\
& & \ \ \ \ \
\left. - {z^2 \over 2}\left( \ln z -1 \right) -{(2-z)^2 \over 2} \ln(2-z)
\right)  \label{eq:notag}
\ena
is a good approximation for total as well as differential rates in the case of
no-tag experiments. With a slight simplification, the anti-tag condition of
TOPAZ \cite {2} excludes events with an electron or a positron with a polar
angle larger than $3.2^{\circ}$ and the above approximation overestimates the rate
of photons \cite {18a}. To illustrate this consider, for example, the reaction
$e^+ e^- \rightarrow e^+ e^- \mu^+ \mu^-$ to the lowest order in QED.
In fig. 6, the differential cross section in $p_T$
of a muon is evaluated with the exact matrix element and compared with the
estimate based on eqs. (\ref{eq:equiv}) and (\ref{eq:notag}): the latter
prediction exceeds the former one by approximately $10\%$ for $p_T =  2\ GeV/c$
and by as much as $100\%$ at the largest $p_T$ values . Under
the experimental conditions of TOPAZ
a much better approximation is based on the following form
\bea
F_{\gamma/ e}(z,E) = {\alpha \over2\pi z} \left( 1+(1-z)^2 \right)
\int^{Q^2_{max}}_{Q^2_{min}} {d Q^2 \over Q^2}
\label{eq:WW}
\ena
appropriate for small angle scattering and which explicitly exhibits the
dependence on the photon propagator. The boundary conditions
$Q^2_{min}$ and $Q^2_{max}$ on the photon virtuality depend on the experimental
triggering conditions. For an anti-tag  condition defined by an angle
$\theta_{max}$ one has $Q^2_{min} =  m_e^2 z^2 / (1-z)$ and
$Q^2_{max} = E^2 \theta^2_{max} (1-z)$  so that the appropriate
Weizs\"acker-Williams spectrum becomes
\bea
F_{\gamma/ e}(z,E) = {\alpha \over \pi z} \left( 1+(1-z)^2 \right)
   \ln{ (1-z) E \theta_{max} \over z\ m_e }
\label{eq:antitag}
\ena
In that case, the approximation overestimates the true result by about roughly
$8\%$ independently of $p_T$ up to $8.5 \ GeV/c$. We shall therefore base our
predictions on eq. (\ref{eq:antitag}) taking into account this
small correction factor. Strictly speaking, this procedure is accurate for the
direct production of jets eq. (\ref{eq:dir}) which are produced via the same
diagrams as the muons in the above example. Because of the delicate
compensation mechanism between eqs.
(\ref{eq:dir})-(\ref{eq:df}) discussed in the previous section, we have to use
the same Weizs\"acker-Williams spectrum for all three types of reactions,
at least for those parts depending on the anomalous structure functions.
This turns out to be justified as can be seen from the following arguments
based on a leading logarithmic analysis. The new feature in the $SF$ and $DF$
components compared to the direct process is the logarithmic dependence of the
anomalous photon structure function. The usual behavior in $\ln (M^2 / Q^2_0)$
is correct when the photon virtuality $Q^2$ is small, say $Q^2 < Q^2_0$, (where
we recall that $Q_0$ is the scale below which the perturbative component
vanishes). When the virtuality of the photon exceeds this bound it is $Q^2$
which acts as the cut-off and the photon structure function behaves then as
$\ln (M^2 / Q^2)$ \cite{22}.
Introducing this factor in eq. (\ref{eq:WW}) and using
the anti-tag condition we find for the effective photon spectrum
\bea
F_{\gamma/ e}(z,E) = {\alpha \over2\pi z} \left( 1+(1-z)^2 \right)
   \left( \ln{ M^2 \over Q^2_0}  \ln{Q^2_{max} \over Q^2_{min}} -
\frac{1}{2}
\ln^2{Q^2_{max} \over Q^2_0} \right)
\label{eq:anotag}
\ena
The first term is the appropriate spectrum for a real photon interacting
through its anomalous component while the second term is the correction factor
induced by the virtuality of the exchanged photon. For the experimental
conditions of TOPAZ this correction never exceeds $2\%$ and we therefore neglect
it altogether.

    It remains finally to discuss the effect of the non-perturbative
component. According to the VDM hypothesis and using the one-pole approximation
the hadronic photon propagator is of the type $m^2/(q^2 (q^2-m^2))$ rather than
the usual $1/q^2$ form. The parameter $m$ is the mass of a typical vector
meson $\rho,\ \omega,\ \phi$ and we simply choose $m^2=.5\ GeV^2$ in our model.
Neglecting a further $q^2$ dependence of the structure function of the hadronic
photon we defined an effective spectrum according to eq. (\ref{eq:WW})
\bea
F_{\gamma/ e}(z,E) &=& {\alpha \over2\pi z} \left( 1+(1-z)^2 \right)
\int^{Q^2_{max}}_{Q^2_{min}} {d Q^2 \over Q^2} \left( {m^2 \over Q^2+m^2}
\right)^2  \nonumber \\
&=& {\alpha \over \pi z} \left( 1+(1-z)^2 \right) \left[
   \ln{ (1-z) E \theta_{max} \over z\ m_e }   \right.\nonumber \\
& &\ \ \ \ \ \ \ \left.
         - \frac{1}{2} \left( \ln\left(1+ {(1-z) E^2 \theta^2_{max} \over m^2 }\right)
       +{(1-z) E^2 \theta^2_{max} \over m^2 +(1-z) E^2 \theta^2_{max}}\right)
\right] 
\label{eq:vdmtag}
\ena
Compared to eq. (\ref{eq:antitag}) appropriate for a point-like photon we find
for the hadronic photon a suppression factor of the order of $10\%-12\%$ in the
kinematic range covered by TOPAZ. It is to be noted that for a no-tag experiment
the suppression factor of the VDM component is even bigger since no large
$\ln(E/m_e)$ is left in the above formula which reduces at very high energy to
\bea
F_{\gamma/ e}(z,E) = {\alpha \over 2 \pi z} \left( 1+(1-z)^2 \right)
  \left[ \ln{ (1-z) m^2 \over z^2\ m^2_e } - 1 \right]
\label{eq:vdmnotag}
\ena

\indent

  We are now in a position to make a comparison with the data. This is shown
in fig. 7. The theoretical predictions are based on different choices
for the VDM input: the top curve is obtained when the standard set of structure
functions of \cite{13} is used while the bottom curve is the result when one
arbitrarily takes the non-perturbative input to be 0 in the standard set. The
first remark to be made is the important role played by the non-perturbative
component at least at the low $p_T$ end of the spectrum.
We have also
shown in the figure, as the intermediate curve, the predictions when one
arbitrarily divides the normalization of the hadronic term by a factor 2,
a solution which is still consistent with the data on $F^\gamma_2$ \cite{13}.
Let us note that for $p_T > 4 \ GeV/c$ the data (with statistical and systematic
errors added linearly) are in very good agreement with
all the theoretical curves while for smaller values the standard set of
structure function overestimates the data and the no-VDM set, although
compatible, is somewhat on the low side.
To further test the sensitivity of the non-perturbative input we have also
tried to modify the shape of the quark
and the gluon distributions in the VDM component keeping the overall
normalization as in the standard set and we have found that the predictions
vary very little, at most by a few $\%$: it thus appears that jet data
in \gamgam collisions
are not very sensitive to the details of the parton distributions due to the
extra convolutions involved in going from \gamgam data to $e^+  e^-$ data.
This is illustrated in fig. 8 where we show the effective $x F^+_{u/e} (x)$
quark and $x F_{g/e}(x)$ gluon momentum weighted
distributions in the electron as "seen" at TOPAZ. These curves are obtained
by taking the convolution of the parton densities in the photon,
at a fixed factorization scale, with the Weizs\"acker-Williams spectrum
appropriate to the experimental triggering conditions. The two sets of curves
in each figure correspond to harder and softer parton distributions,
than the standard ones, in the hadronic component of the photon. We choose
\bea
x v (x,2 GeV^2) &\simeq& (1-x)^{.85}                            \nonumber \\
x G (x,2 GeV^2) &\simeq& (1-x)^{1}
\ena
for the hard set and
\bea
x v (x,2 GeV^2) &\simeq& (1-x)^{2}                              \nonumber \\
x G (x,2 GeV^2) &\simeq& (1-x)^{3}
\ena
for the soft set. In each case the normalization factors are appropriately chosen
to satisfy the known sum rules. We recall that the standard parametrization,
as fitted to the pion induced hard cross sections \cite {26}, is
\bea
x v (x,2 GeV^2) &\simeq& (1-x)^{.85}                            \nonumber \\
x G (x,2 GeV^2) &\simeq& (1-x)^{1.94}
\ena
We note that the two extreme sets are very similar in the region $x \simeq .2$.
Since the effective $x$ probed by TOPAZ is estimated to be $x \simeq .15$ at
$p_T = 3\ GeV/c$ and $x \simeq .25$ at $p_T = 7\ GeV/c$, it is not surprising
that the data are rather insensitive to changes in the shape of the hadronic
component.

        It may be interesting to discuss at this point the role of the
next-to-leading logarithm (NTL) corrections. The comparison with the leading
logarithm (LL) results is made using the same scale $\mu=M=p_T$
and the standard
set of structure functions. Thus we discuss only the size of the correction
factors $K^D$, $K^{SF}$ and $K^{DF}$ of eq. (\ref{eq:compens}). Globally we
find that the NTL corrections increase the cross section by $25\%$ at
$p_T = 3\ GeV/c$ and only $3\%$ at $p_T = 7.5\ GeV/c$. The higher order
terms affect the components rather differently since the direct cross section
is reduced by $15\%$ throughout the $p_T$ range covered by the data, the SF
one is left approximately unchanged (a few $\%$ variation) and the DF one
increases by $70\%$. The net result of this pattern of corrections is that
the direct cross section is reduced from $35\%$ to $24\%$
of the total jet cross section at $p_T = 3.12\ GeV/c$
and from $51\%$ to $42\%$ at $p_T = 7.5\ GeV/c$. Let us note
finally the relative importance of the DF term since it accounts
for $55\%$ of the total at $p_T = 3.12 \ GeV/c$ and still $31\%$ at
$p_T = 7.5\ GeV/c$ Needless to say that all these
figures would be affected had we used a different jet definition.

  The discrepancy between theory and experiment at low $p_T$ may, after all,
not be too surprising. Because of the smallness of the scales involved the
perturbative regime may not have set in. In other words, non-perturbative
intrinsic $p_T$ effects, which have been ignored in our calculation, could
somewhat affect the theoretical predictions.
More probably however, the relevant point is the neglect of the charm quark
mass in the hard partonic cross sections as well as in the evolution of the
structure functions except for the charm threshold at $M^2 = 2 \ GeV^2$.
An estimate of this effect may be attempted by comparing, in the leading
logarithmic approximation, the direct cross section using four massless quarks
with that assuming three massless flavors and a charm quark of mass
$1.5 \ GeV/c^2$.
In the latter case, the cross section is reduced by $14\%$ at $p_T = 3 \ GeV/c$,
$5\%$ at $p_T = 5 \ GeV/c$ and only $2\%$ at $p_T = 7 \ GeV/c$. Such factors
when applied to our predictions would put the two VDM curves in perfect
agreement with the data and disfavor the no-VDM predictions. Of course such
corrections factors are only indicative since the physics of jet production
is much more complex than the simple direct production specially at small
transverse momentum.

  In fig. 9 we show the predictions for the jet $p_T$ spectrum at LEP 200 with a
no-tag condition and in fig. 10 those at a collider with an energy
\rsee $=\ 1\ TeV$. Despite the large scales involved it is interesting to
remark that the non-perturbative term still gives a large contribution
up to rather high 
$p_T$  values. This is related to the fact that higher energy experiments
tend to probe smaller $x$ values of the photon structure function where the
hadronic component is important.

\section {Conclusions}

\indent

    Despite the rather large intrinsic scale uncertainties involved in the
theoretical predictions for single jet production in \gamgam collisions,
useful phenomenology can be done. Good agreement is obtained with TOPAZ
data for $p_T > 4\ GeV/c$. The rather large error bars (both experimental
and theoretical) do not allow, for the moment, to distinguish the predictions
based on different sets of structure functions: both the set without a
non-perturbative component
and the set with the theoretically expected VDM input (standard set) are
compatible with the data. For $p_T < 4\ GeV/c$, the charm mass neglected in the
calculation, plays an important role. Correcting semi-quantitatively for this
mass effect we find the standard set to be in agreement with the data while the
other one gives too small a cross section. Another feature of our work is the
importance of higher order corrections to processes where the photon
couples through its structure function. This means that at low jet $p_T$ values
the direct process is relatively reduced and therefore
\gamgam collisions tend to look like hadron-hadron collisions with some jet
activity in the forward and backward direction.

        We find it remarkable that QCD is able to correlate in a rather
consistent way the measurements on $F_2^\gamma$ to the rate of jet production
in \gamgam collisions. The agreement is even more remarkable if one remembers
that the mass scale involved, typically the jet transverse momentum, are
rather small. We eagerly wait for more precise data to put the theory to an
even more stringent test.

        As seen in our analysis, some complications in the comparison between
theory and experiment arose from the virtuality of the photons: theoretical
predictions are made for real photons while experiments deal with off-shell
photons. This problem can be handled in the framework of the Vector Dominance
Model. It would make the comparison between theory and experiment more direct
if experimental triggering conditions could be such as to minimize the
virtuality of the photons and this wish also applies to measurements of
$F_2^{\gamma}$ as well as of jet production in \gamgam collisions.

\newpage

{\Large{\bf Acknowledgements}}

\indent

   We thank H. Hayashii for many informative and helpful discussions as
well as N. Nakazawa for early collaboration on this work. We are indebted
to CNRS-IN2P3(France) and Monbusho(Japan) for financial support to our
collaboration.


\newpage

{\Large{\bf Figure Captions}}

\vspace{9 mm}

\begin{tabular}{ll}
Fig. 1 & Direct processes in $\gamma-\gamma$ collisions. \\
Fig. 2 & SF type processes in $\gamma-\gamma$ collisions. \\
Fig. 3 & DF type processes in $\gamma-\gamma$ collisions. \\
Fig. 4 & Variation of the cross section  $\dsigee$ at \rsee$= 58\ GeV$,\\
 & $p_T = 5.24\ GeV/c$, $\eta=0$, with the factorization scale $M$ and \\
 & the renormalization scale $\mu$: a) leading logarithm predictions; \\
 & b) next-to-leading logarithm predictions.
Note that in order to fully \\
 & display the shape of the surface we rotated the figure by $90^0$ compared \\
 & a).\\
Fig. 5 & The same as fig. 1 for \rsee$= 1\ TeV$, $ p_T = 50\ GeV/c$. \\
Fig. 6 & Study of the equivalent photon approximation. $p_T$ spectrum of the\\
 & muon in the process $e^+ e^- \rightarrow e^+ e^- \mu^+ \mu^-$ at
\rsee$= 58\ GeV$. The \\
 & data points are the result of the exact calculation, the top line is the \\
 & no-tag prediction using the Weizs\"acker-Williams approximation eq.
(\ref{eq:notag}) \\
 & and the bottom line is from eq. (\ref{eq:antitag}) with
$\theta_{max}=3.2^\circ$.\\
Fig. 7 & TOPAZ data on inclusive jet production and theoretical predictions
for  \\
 & $\int^{.7}_{.7} d\eta {d \sigma^{e^+e^- \rightarrow jet}
\over {d p_T d \eta}}$. The small error bars are statistical(a) and the large ones \\
 & are statistical and systematic errors added linearly(b). The top curve is \\
 & the theoretical prediction based on the standard photon structure functions,\\
 & the middle one is based on structure functions with half the VDM input, \\
 & and the lower one is based on the perturbative component only. \\
Fig. 8 & Distributions of a) quarks and b) gluons in the electron under TOPAZ \\
 & triggering conditions. The solid curve corresponds to the hard input and \\
 & the dashed one to the soft input. The factorization scale is
$M^2= 25\ GeV^2$.\\
Fig. 9 & Single jet production, via the two-photon mechanism, at LEP 200 for a
   \\
 & no-tag experiment and $\eta= 0$. The top and bottom curves have the same \\
 & meaning as in fig. 7. \\
Fig. 10 & Same as in fig. 9 at \rsee$=1\ TeV$. \\

\end{tabular}


\begin{thebibliography}{99}
\bibitem{18a}   In fact TOPAZ accepts events at larger polar angles only when
                the quasi-real photon carries more than 75\% of the electron
                momentum. Such an effect is included in our predictions but
                it hardly modifies the jet spectrum in the measured transverse
                momentum range.
\end{thebibliography}
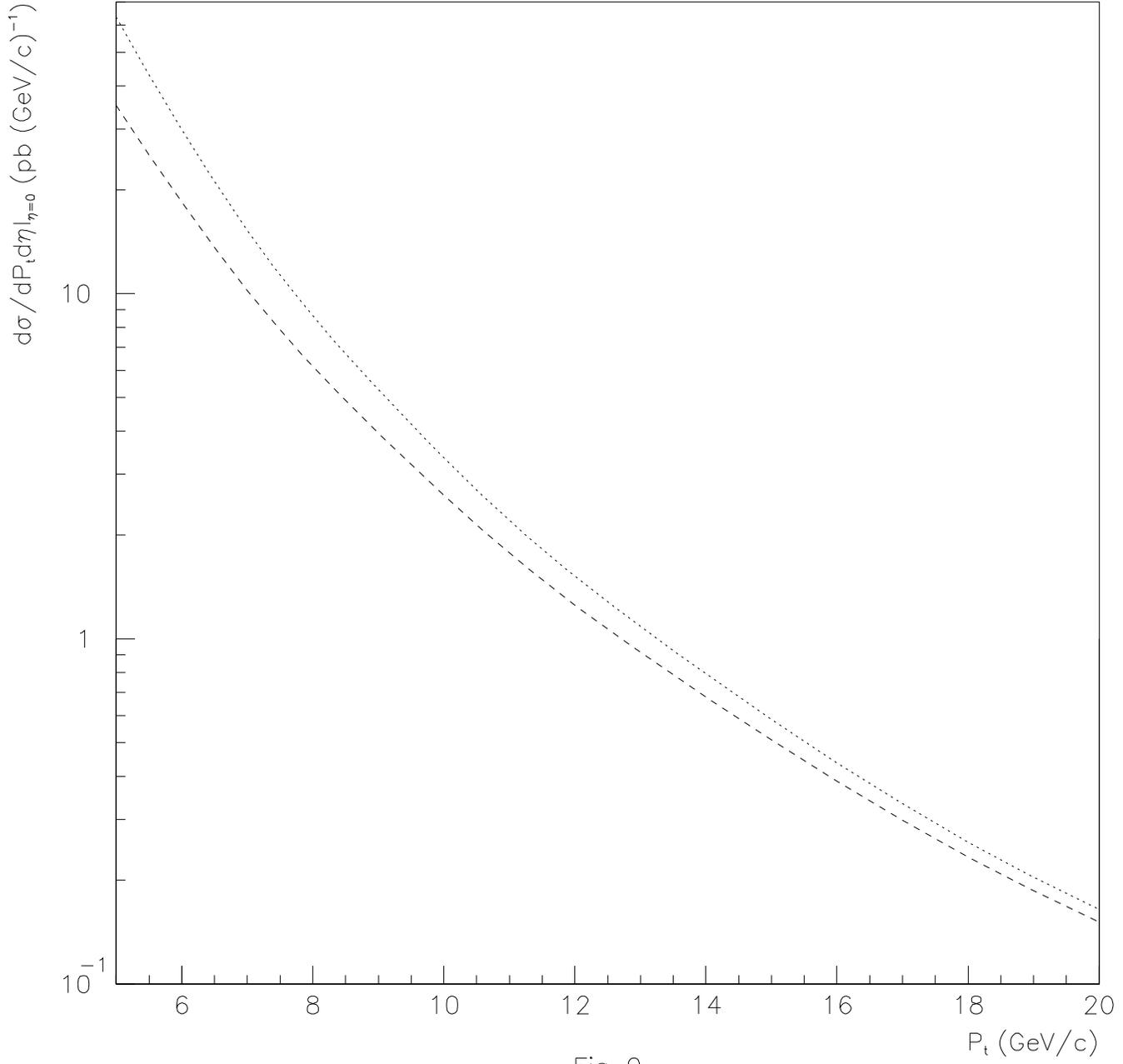
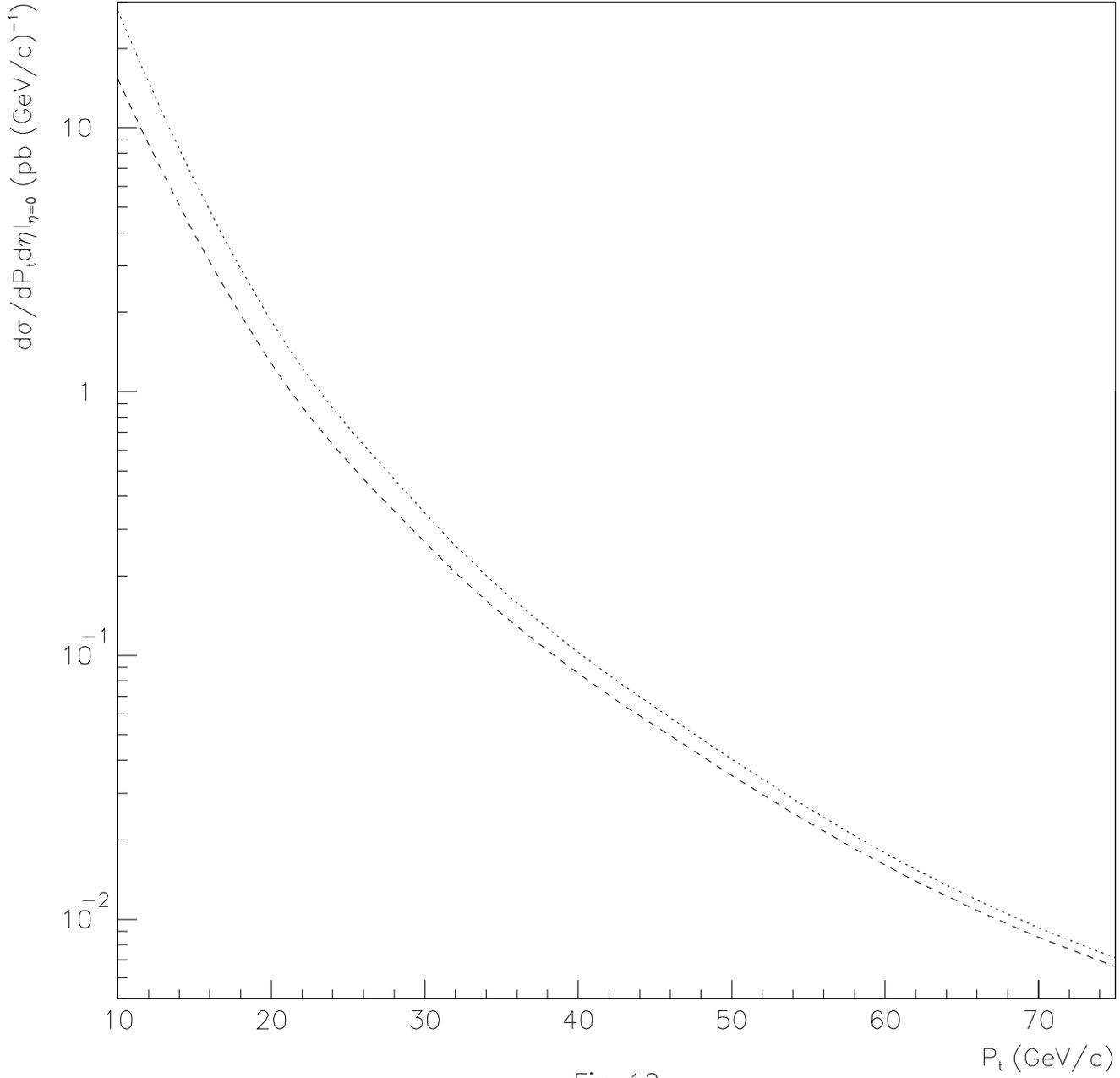
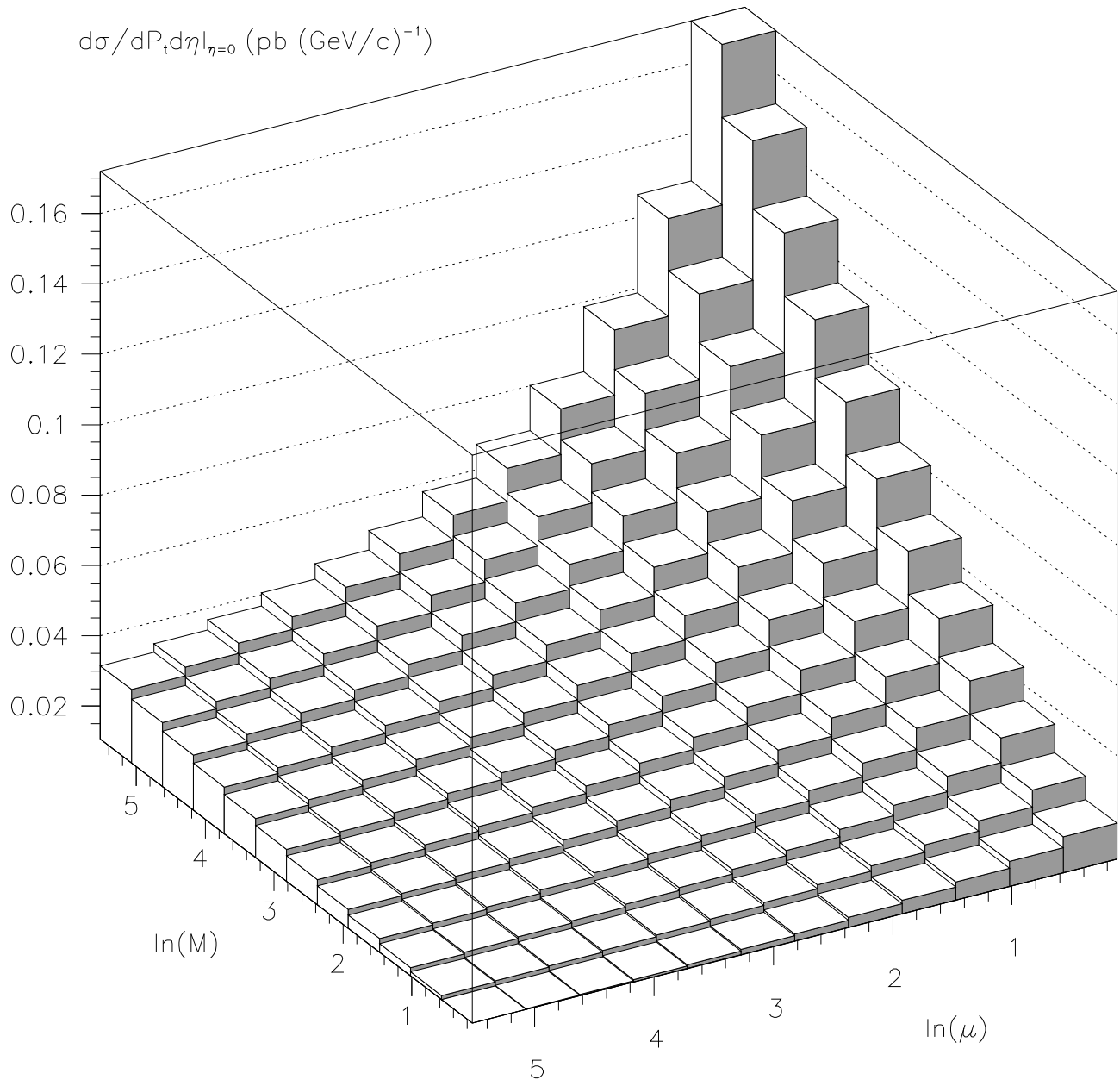
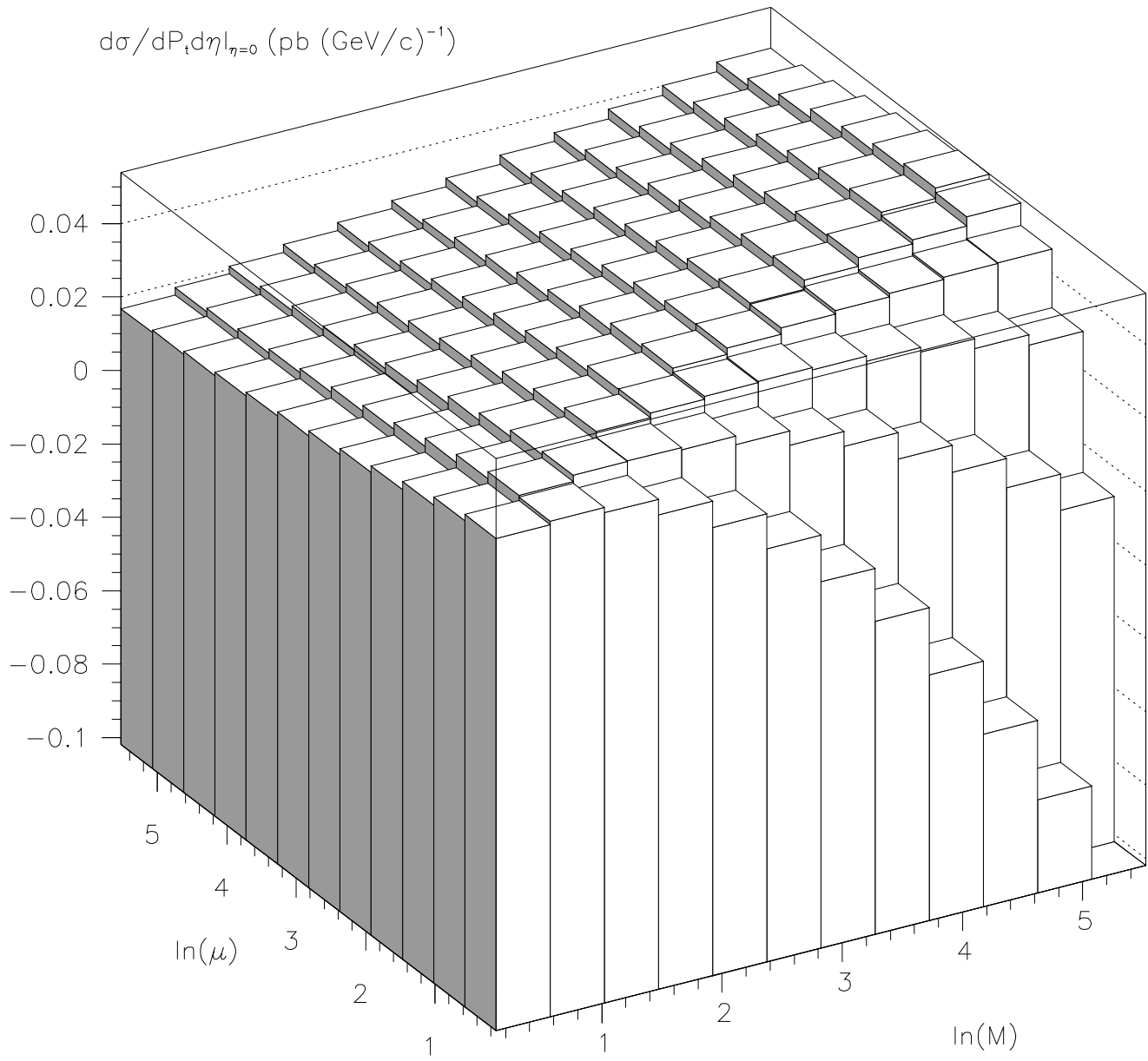
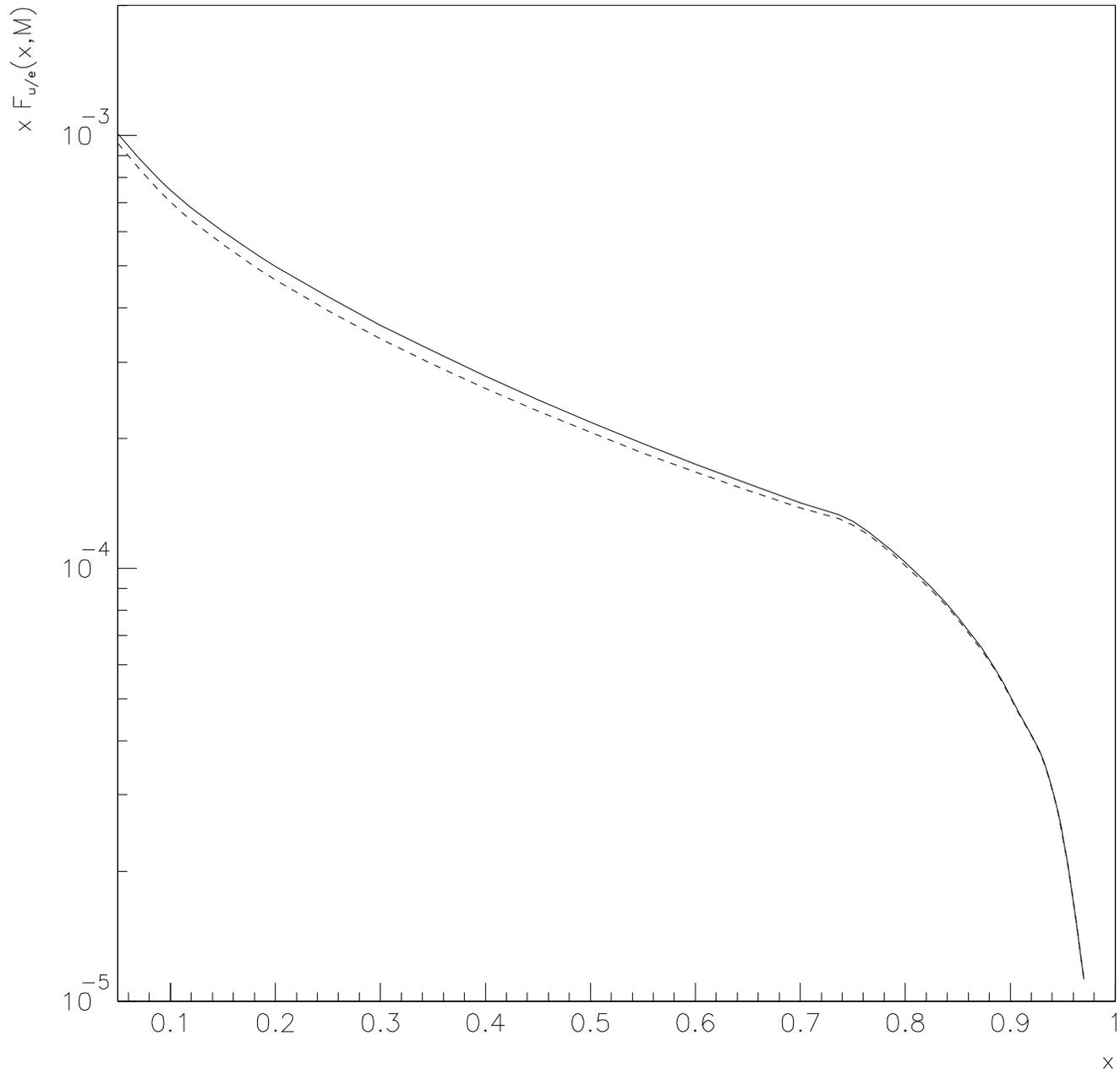
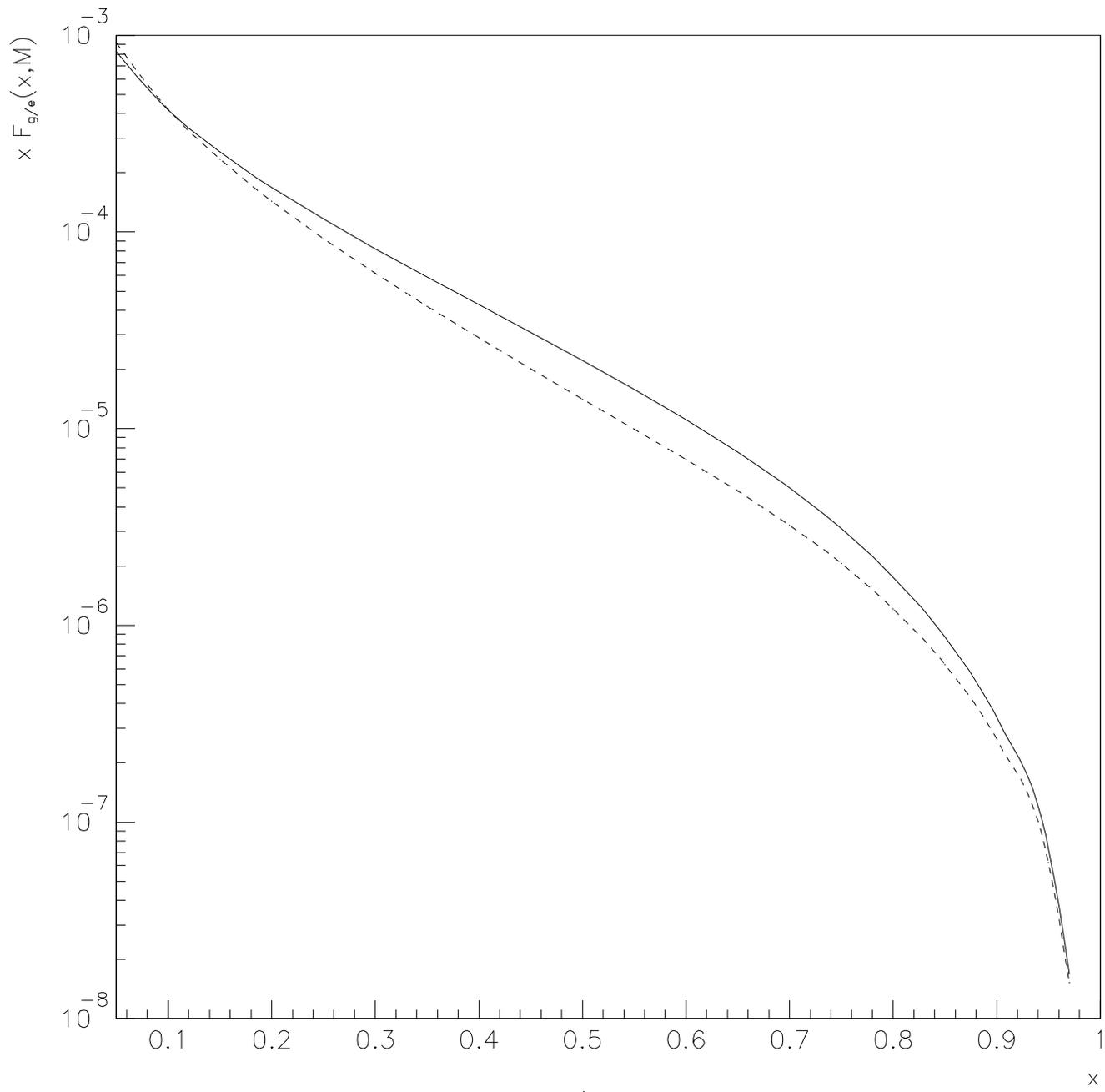
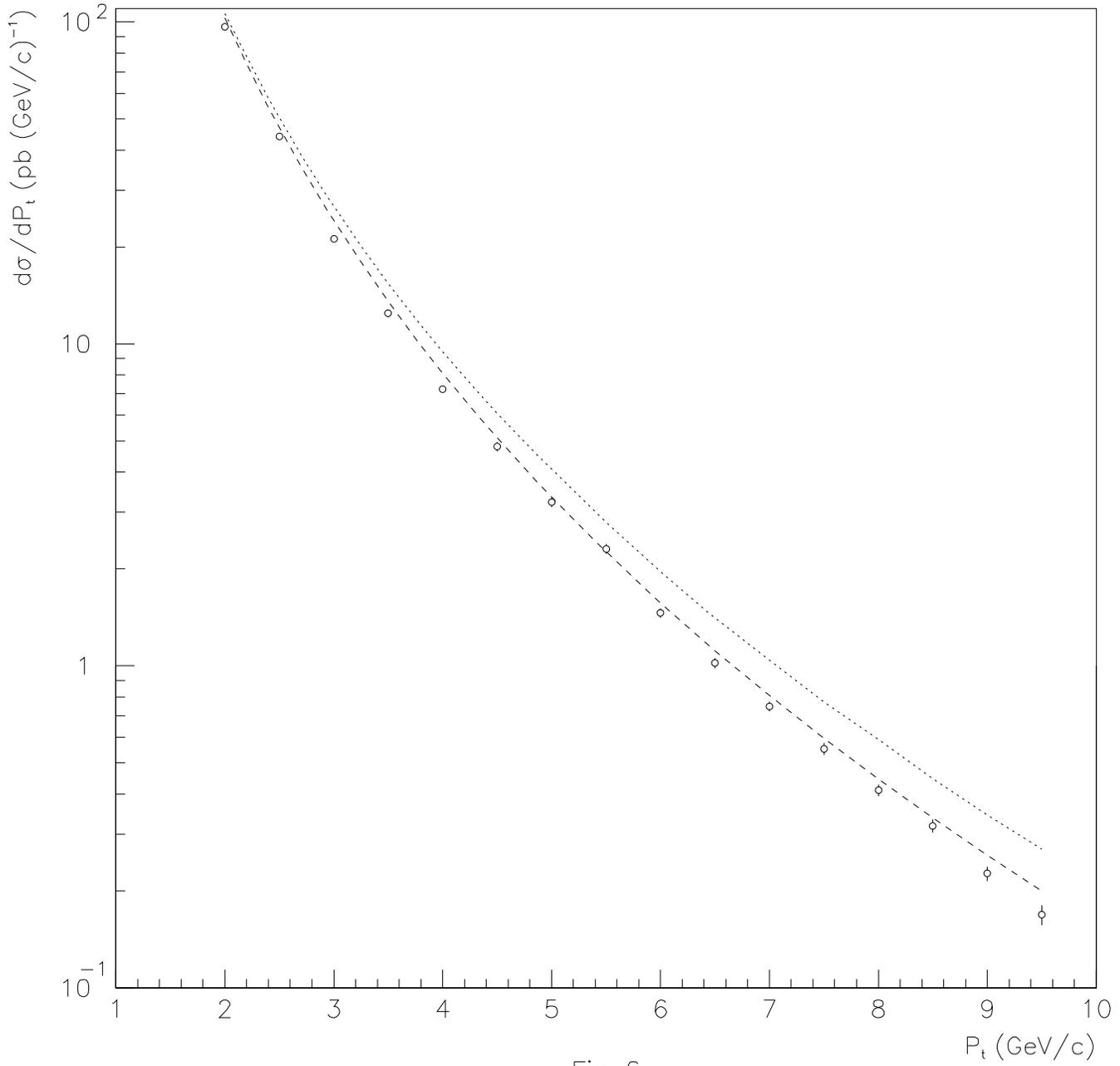
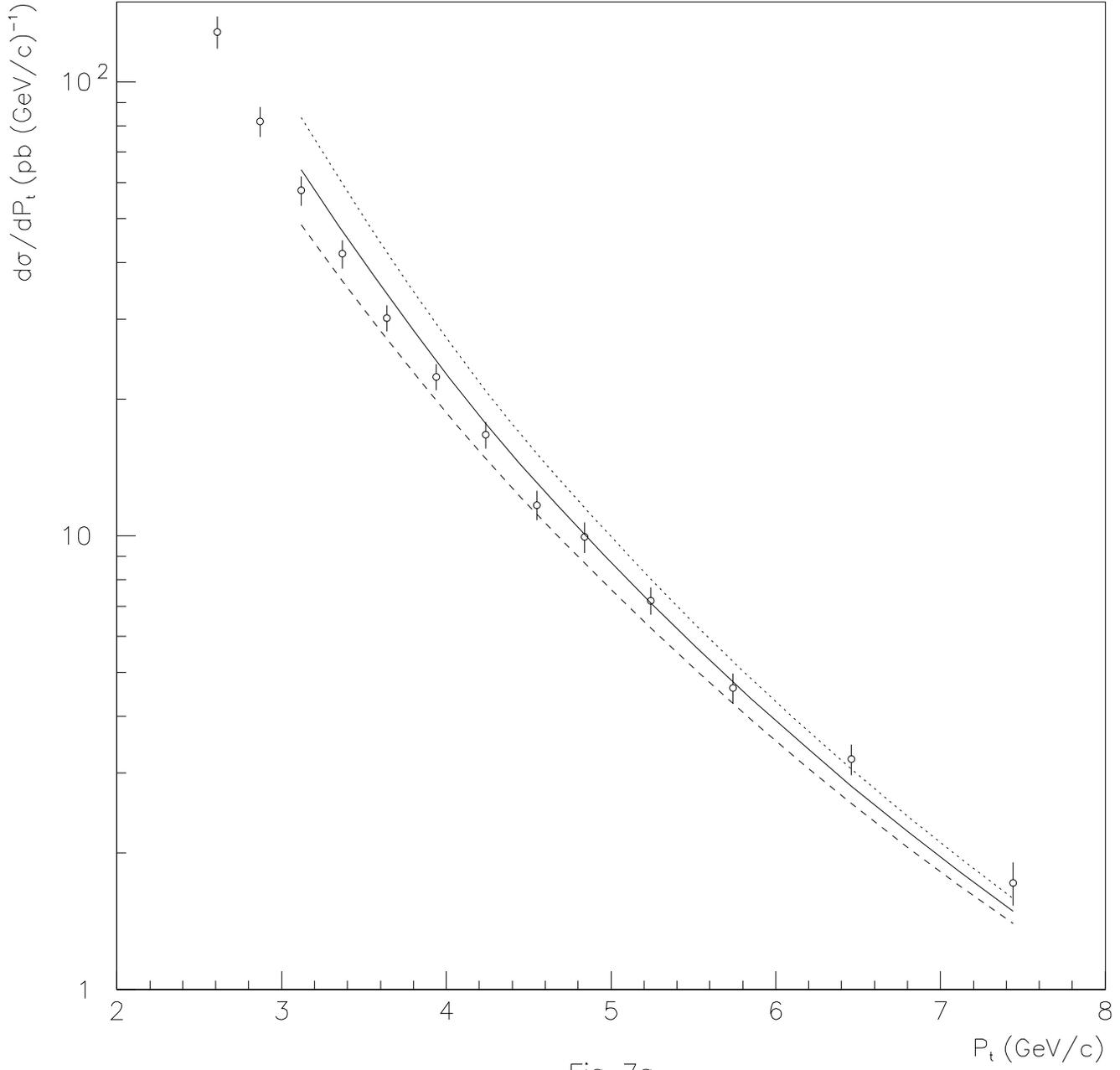
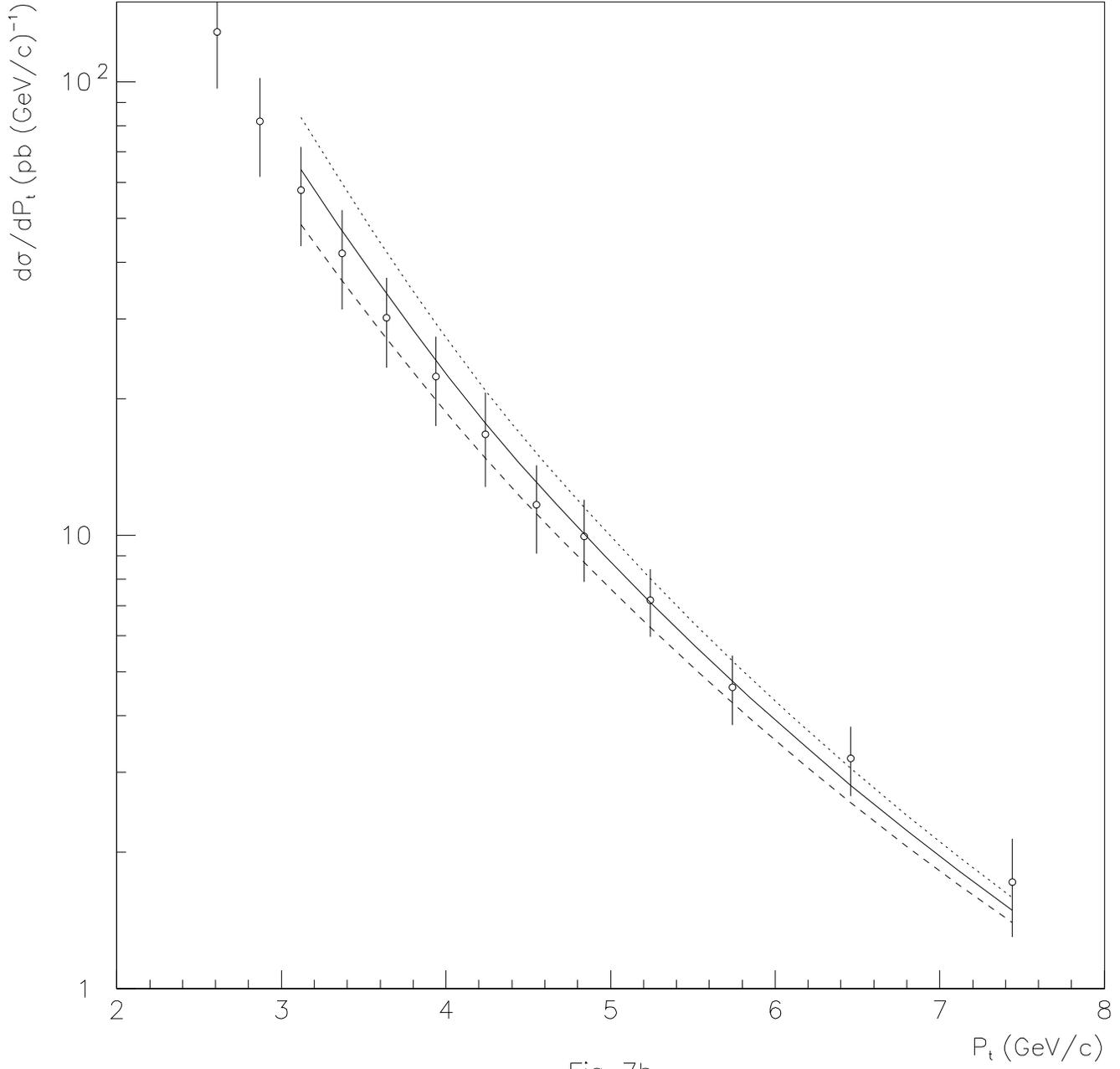
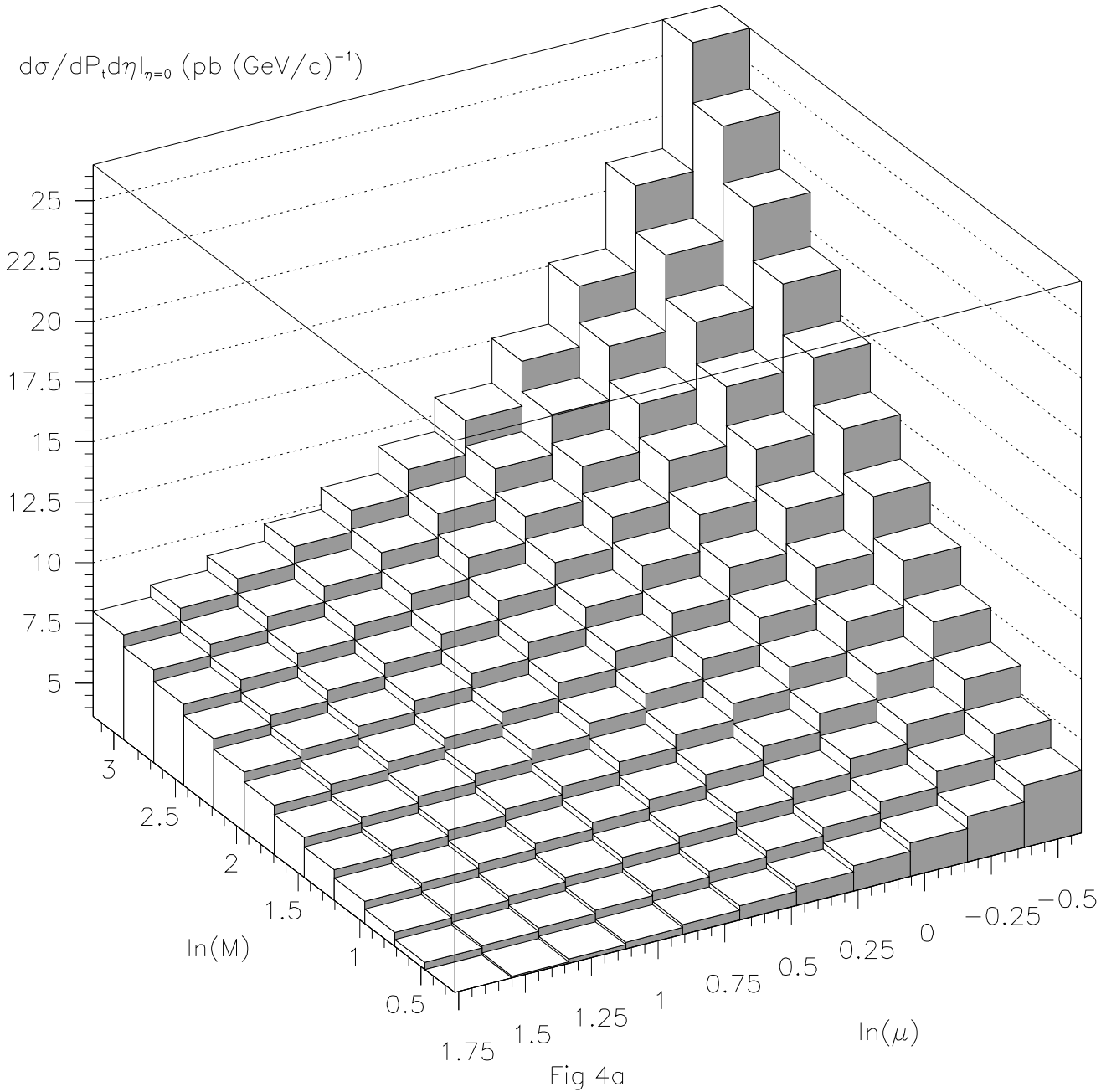
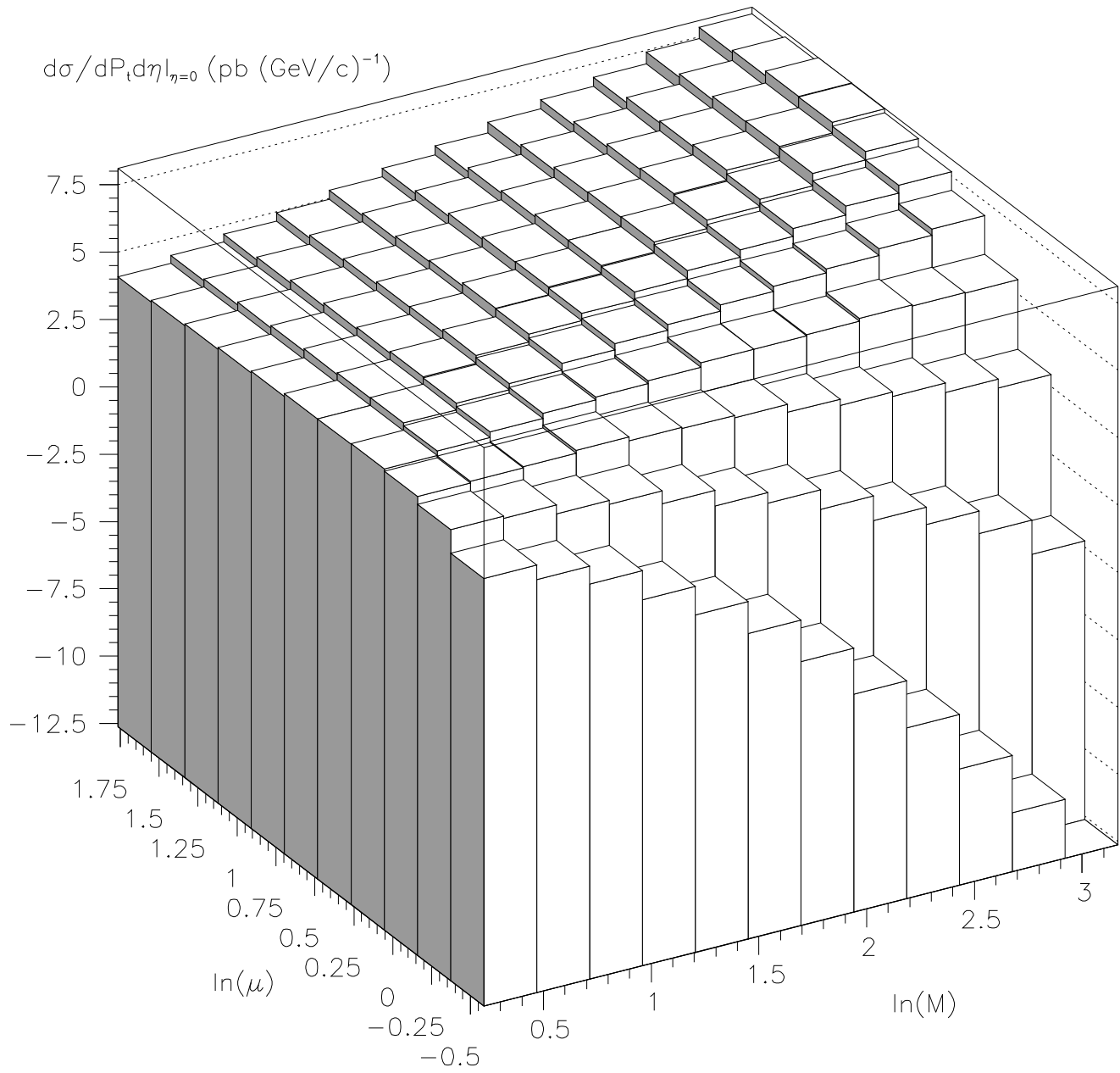
\end{document}